\def\rn{\noindent\parshape 2 0truecm 8.5truecm 0.3truecm 8.2truecm}
\def\rn{}
\def\nn#1 #2{#2. #1}				
\def\nnn#1 #2 #3{#2. #3. #1}			
\def\nnnn#1 #2 #3 #4{#2. #3. #4 #1}		
\def\nnnnn#1 #2 #3 #4 #5{#2. #3. #4 #5. #1}	
\def\dualand{ and\hbox{ }}				
\def\multiand{, and\hbox{ }}				
\def\rf#1;#2;#3;#4;#5 {{\frenchspacing\par\rn#1, #3 {\bf #4}, #5 (#2). \par}}
\def\rg#1;#2;#3;#4;#5;#6 {{\frenchspacing\par\rn#1, #3 {\bf #4}, #5 (#2). \par}}
\def\rfbook#1;#2;#3;#4;#5 {{\frenchspacing\par\rn#1, {\it #3} (#5, #4, #2).\par}}
\def\rfprep#1;#2;#3 {{\par\frenchspacing\rn#1, #3 (#2).\par}}

\def\Mpc{{\rm Mpc}}

\def\eV{{\rm eV}}

\def\etal{{\frenchspacing\it et al.}}
\def\ie{{\frenchspacing\it i.e.}}
\def\eg{{\frenchspacing\it e.g.}}
\def\etc{{\frenchspacing\it etc.}}

\def\beq#1{\begin{equation}\label{#1}}
\def\eeq{\end{equation}}
\def\beqa#1{\begin{eqnarray}\label{#1}}
\def\eeqa{\end{eqnarray}}
\def\eq#1{equation~(\ref{#1})}

\def\eqn#1{~(\ref{#1})}

\def\fig#1{Figure~\ref{#1}}
\def\Fig#1{Figure~\ref{#1}}

\def\sec#1{Section~\ref{#1}}

\def\spose#1{\hbox to 0pt{#1\hss}}
\def\simlt{\mathrel{\spose{\lower 3pt\hbox{$\mathchar"218$}}
     \raise 2.0pt\hbox{$\mathchar"13C$}}}
\def\simgt{\mathrel{\spose{\lower 3pt\hbox{$\mathchar"218$}}
     \raise 2.0pt\hbox{$\mathchar"13E$}}}
\def\simpropto{\mathrel{\spose{\lower 3pt\hbox{$\mathchar"218$}}
     \raise 2.0pt\hbox{$\propto$}}}

\def\ed{\end{document}}

\def\Ob{\Omega_{\rm b}}
\def\Oc{\Omega_{\rm cdm}}
\def\Od{\Omega_{\rm dm}}
\def\Ok{\Omega_{\rm k}}
\def\Ol{\Omega_\Lambda}
\def\Om{\Omega_{\rm m}}
\def\On{\Omega_\nu}
\def\ob{\omega_{\rm b}}
\def\ocdm{\omega_{\rm cdm}}
\def\od{\omega_{\rm dm}}

\def\om{\omega_{\rm m}}
\def\on{\omega_\nu}
\def\fn{f_\nu}
\def\Cl{C_\l}

\def\ns{n_s}
\def\nt{n_t}
\def\As{A_s}
\def\At{A_t}
\def\dA{d_{\rm lss}}

\def\zion{z_{ion}}

\def\L{{\cal L}}

\def\p{{\bf p}}

\def\l{\ell}

\def\Cl{C_\ell}
\def\Clo{C_\ell^{\rm low}}
\def\Chi{C_\ell^{\rm high}}

\def\shor{d_{\rm shor}}

\documentstyle[prd,aps,epsf]{revtex}
\begin{document}
\twocolumn[\hsize\textwidth\columnwidth\hsize\csname@twocolumnfalse\endcsname



\title{Towards a refined cosmic concordance model: joint 11-parameter 
constraints from CMB and large-scale structure}

\author{Max Tegmark}

\address{Dept. of Physics, Univ. of Pennsylvania, 
Philadelphia, PA 19104; max@physics.upenn.edu}

\author{Matias Zaldarriaga}

\address{Institute for Advanced Study, Princeton, 
NJ 08540; matiasz@ias.edu}

\author{Andrew J. S. Hamilton}

\address{JILA and Dept.\ of Astrophysical and Planetary Sciences,
Box 440, Univ. of Colorado,\\
Boulder, CO 80309; Andrew.Hamilton@colorado.edu}

\date{\today. Accepted for publication in Phys. Rev. D.}

\maketitle

\begin{abstract}

We present a method for calculating large numbers of power
spectra $C_\l$ and $P(k)$ that accelerates CMBfast by a factor around 
$10^3$ without appreciable loss of accuracy,
then apply it to constrain 11 cosmological 
parameters from current Cosmic Microwave Background (CMB)
and Large Scale Structure (LSS) data. 
While the CMB alone still suffers from several degeneracies,
allowing, \eg, closed models
with strong tilt and tensor contributions, 
the shape of the real space power spectrum of galaxies
from the {\it IRAS} Point Source Catalogue Redshift (PSCz) survey
breaks these degeneracies and 
helps place strong constraints on most parameters.
At 95\% confidence, the combined CMB and LSS data imply
a baryon density $0.020 < \ob < 0.037$,
dark matter density $0.10<\od< 0.32$
with a neutrino fraction $\fn < 38\%$,  
vacuum density $\Ol < 0.76$, 
curvature $-0.19 < \Ok < 0.10$, 
scalar tilt $0.86 < \ns < 1.16$, and
reionization optical depth $\tau < 0.44$.
These joint constraints are quite robust, changing little
when we impose priors on
the Hubble parameter, tilt, flatness, gravity waves or reionization.
Adding nucleosynthesis and neutrino priors on the other hand
tightens constraints considerably,
requiring $\Ol>0.49$ and a red-tilt, $\ns < 1$.

The analysis allows a number of consistency tests to be made, 
all of which pass.
At the 95\% level, 
the flat scalar ``concordance model'' with
$\Ol=0.62$,
$\od=0.13$,
$\ob=0.02$,
$\fn\sim 0$,
$\ns=0.9$,
$\tau=0.1$,
$h=0.63$
is consistent with the CMB and LSS data considered here,
with big bang nucleosynthesis,
cluster baryon fractions and cluster abundance.
The inferred PSCz bias $b\sim 1.2$ agrees with the value
estimated independently from redshift space distortions.
The inferred cosmological constant value agrees with the one 
derived independently from SN 1a studies.
Cosmology seems to be on the right track!
\end{abstract}



] 


\begin{figure}[tb] 
\centerline{\epsfxsize=8.5cm\epsffile{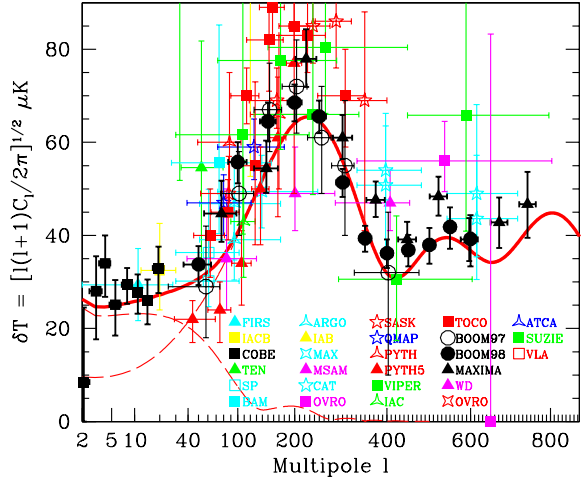}}
\smallskip
\caption{\label{DegenFig}\footnotesize%
The importance of including non-CMB information.
The best fit to all current CMB data (solid curve) is still a crazy
tilted model ($\ns=1.4$) for a closed universe
($\Omega_{\rm tot}=1.3$) with 
the 
COBE signal explained almost entirely
with tensor fluctuations.
The dashed curves show the tensor and scalar contributions.
}
\end{figure}

\section{INTRODUCTION}

The cosmic microwave background (CMB) is dramatically improving our
knowledge of cosmological parameters
\cite{Lange00,boompa,Bambi00,observables,Jaffe00,Kinney00},
although 
as can be seen in \fig{DegenFig}, the constraints from
CMB alone are weaker than is sometimes claimed.
The figure shows that the
adiabatic inflationary model that fits the CMB data best  
is still a crazy
one with extreme scalar tilt $\ns=1.4$, curvature
($\Omega_{\rm tot}=1.3$) and the COBE signal explained almost entirely
with tensor fluctuations (gravity waves).

Indeed, CMB data have now become so sensitive that
the key issue in cosmological parameter determination
is not always the accuracy with which CMB power spectrum features
(such as the position of the first peak) can be measured, but often
what prior information  
is used or assumed (e.g., that there are no tensor fluctuations). 
A range of priors were explored in
recent studies
\cite{Lange00,boompa,Bambi00,observables,Jaffe00,Kinney00},
including assumptions about reionization and 
gravity waves and
constraints from nucleosynthesis, supernovae, large-scale structure and
the Hubble constant.
In particular, 
two extensive multiparameter analyses \cite{Lange00,Jaffe00}
included large-scale structure (LSS) information 
as quantified by a normalization 
and the so-called ``shape parameter'' $\Gamma$ (which slides
a fixed transfer function sideways) from
galaxy power spectrum measurements and cluster abundances,
extending earlier CMB+LSS work
\cite{Kofman93,WhiteBaryons,Liddle96,Bunn97,Gawiser98,Webster98,Bond98,CosmicTriangle,Novosyadlyj00,Bridle99,Bridle00}.

While much of the information in the galaxy power spectrum is indeed
encapsulated in a horizontal and a vertical offset,
all of it is clearly not.
It should therefore be possible to do still better by 
fitting directly to the LSS data, 
explicitly including the 
way in which each of the cosmological parameters affect this
curve, just as is presently done for the CMB.
This is the goal of the present paper.
Now is a particularly exciting time to start doing this,
since projects like the 2dF Survey and the 
Sloan Digital Sky Survey will soon produce 
dramatic improvements in LSS data quality.

\begin{figure}[tb] 
\centerline{\epsfxsize=8.5cm\epsffile{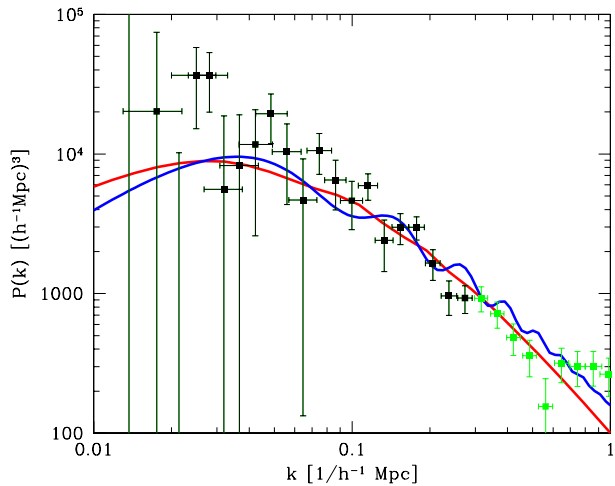}}
\caption{\label{PSCzFig}\footnotesize%
Why LSS data adds information to the CMB.
The wiggly curve corresponds the best fit model from CMB alone
that was shown in Figure 1, normalized on small scales
(it has bias $b=0.5$).
The straighter curve shows Prior P5 from
\protect\cite{Jaffe00}
--- the difficulty in matching the largest scales illustrates that
there is already more information in the curve than a
normalization and a shape parameter.
Of the PSCz measurements shown,
we opt to use only those in the fairly linear 
regime $k<0.3 \, h\,\Mpc^{-1}$ (black), discarding the rest.
}
\end{figure}

The LSS data used in this paper is the linear real space power spectrum
of the {\it IRAS} Point Source Catalogue Redshift Survey \cite{Saunders00}
(PSCz)
as measured by \cite{pscz}
and as shown in Figure~\ref{PSCzFig}.
The PSCz survey contains
redshifts for
14{,}677
galaxies covering 84\%
of the sky to a usable depth of about $400 \, h^{-1}\Mpc$.
Although other large 
data sets such as the 
Las Campanas Redshift Survey 
\cite{Shectman96}
and the CfA/SSRS UZC redshift 
survey 
\cite{Falco99}
have comparable numbers of galaxies,
the large volume of the PSCz,
along with the careful attention paid by its
authors to uniformity of selection,
makes PSCz the most powerful publicly available probe of LSS
at large, linear scales.

Afficionados may notice that the error bars on the PSCz power spectrum
in Figure~\ref{PSCzFig} appear somewhat larger than some other
published measurements.
This is because the measurements have been decorrelated
\cite{galpower2}
so that each plotted point represents an essentially independent piece of
information.
Having uncorrelated data points,
or equivalently a full covariance matrix,
is prerequisite for a reliable likelihood analysis.

A longstanding obstacle to interpreting LSS measurements
is the thorny issue of galaxy-to-mass bias.
Local bias models predict that the bias factor should be constant
at large, linear scales
\cite{Coles93,Fry93,Scherrer98,Coles99},
and $N$-body experiments tend to confirm this notion
\cite{Kravtsov99,Colin99,Narayanan00,Benson00}.
The simple situation at linear scales contrasts with the nonlinear regime,
where the afore-referenced $N$-body experiments suggest that there is likely
to be substantial scale-dependent bias.
For this reason,
we confine the analysis of the present paper to the linear regime,
$k < 0.3 \, h \, \Mpc^{-1}$. We return to this issue below.

We will investigate how CMB and LSS 
constrain cosmological parameters, both jointly and in separate ways 
that allow consistency checks to be made. In order to be able to study
the effects of prior assumptions, this
forces us 
to work in an 11-dimensional parameter space. To make this feasible
in practice, we first need to develop and test a method for 
computing theoretical power spectra $C_l$ and
$P(k)$ accurately and rapidly.

The rest of this paper is organized as follows. 
In \sec{MethodSec}, we summarize our
methods for computing CMB and LSS
power spectra, saving the implementation details and
tests of their accuracy for Appendices A, B and C.
We present our constraints on cosmological parameters in 
\sec{ResultsSec} and discuss our conclusions in \sec{ConclusionsSec}.

\section{METHOD}

\label{MethodSec}

Our method is based on the one described in \cite{10par},
but with a number of extensions and improvements as detailed 
in Appendices A, B and C.
It consists of the following steps:
\begin{enumerate}
\item Compute power spectra $C_\l$ and $P(k)$ for a grid
of models in our 11-dimensional parameter space.
\item Compute a likelihood for each model that quantifies how well it fits the
data.
\item Perform 11-dimensional interpolation and marginalize to obtain
constraints on individual parameters and parameter pairs.
\end{enumerate}
Our main improvement over \cite{10par} is in step 1.
As described in Appendix B, we enhance the technique for accelerated 
CMB power
spectrum calculation so that it becomes essentially as 
accurate as CMBfast itself,
but about $10^3$ times faster.
We also add a simple but accurate technique to compute the grid 
of matter power spectra rapidly as described in Appendix C.
In addition, we improve the choice of parameters and gridding from \cite{10par}
as detailed in Appendix A.
For the reasons given there, we use the 11 parameters 
\beq{pEq}
\p\equiv(\tau,\Ok,\Ol,\od,\ob,\fn,\ns,\nt,\As,\At,b).
\eeq
These are the reionization optical depth $\tau$, 
the primordial amplitudes $\As$, $\At$ and tilts $\ns$, $\nt$ 
of scalar and tensor fluctuations, 
a bias parameter $b$ defined as the ratio between rms 
galaxy fluctuations and rms matter fluctuations on 
large scales,
and five parameters specifying the cosmic matter budget.
The various contributions $\Omega_i$ to critical density are for
curvature $\Ok$, vacuum energy $\Ol$, cold dark matter $\Oc$, 
hot dark matter (neutrinos) $\On$ and baryons $\Ob$.
The quantities
$\ob\equiv h^2\Ob$ and
$\od\equiv h^2\Od$ correspond to 
the physical densities of baryons
and total (cold + hot) dark matter 
($\Od\equiv\Oc+\On$), and $\fn\equiv\On/\Od$ is the fraction
of the dark matter that is hot.
We assume that the bias $b$ is constant on large scales 
but make no assumptions about its value, 
and therefore marginalize (minimize) over this parameter
before quoting constraints on the other ten.

\section{RESULTS}

\label{ResultsSec}

\begin{figure}[tb] 
\centerline{\epsfxsize=8.5cm\epsffile{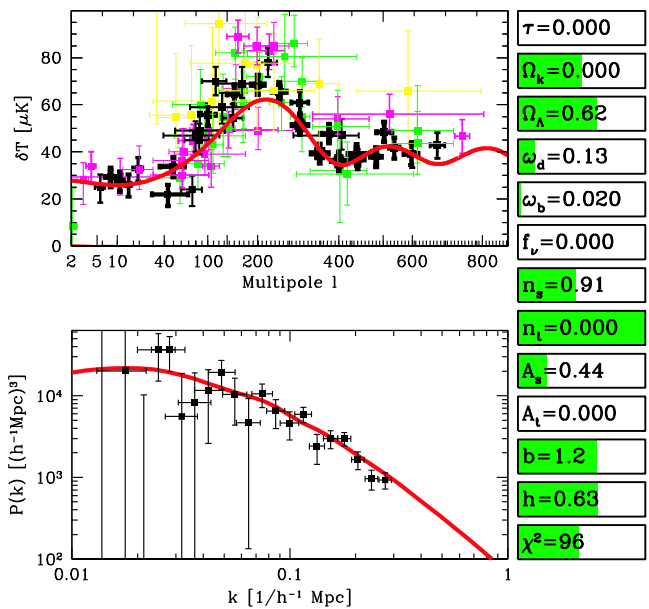}}
\bigskip
\caption{\label{CPfig}\footnotesize%
The CMB and LSS power spectra for the ``concordance'' model from Table 1.
Animated versions of this figure, where the effect of changing one parameter
at a time can be viewed, are available at 
www.hep.upenn.edu/$\sim$max/concordance.html.
The plotted model has bias
$b=1.24$,
redshift distortion parameter $\beta=0.47$
and hubble parameter $h=0.63$.
It provides an acceptable fit to all our data, with 
$\chi^2\approx 96$ for about $109-11=98$ degrees of freedom.
}
\end{figure}

\subsection{Basic results}

\begin{figure}[tb] 
\centerline{\epsfxsize=8.5cm\epsffile{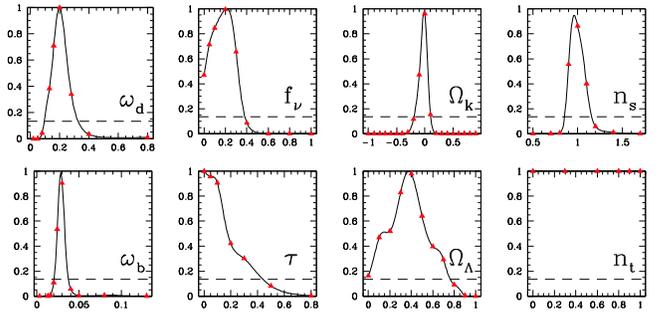}}
\bigskip
\caption{\label{1DnoFig}\footnotesize%
Constraints on individual parameters using only CMB and LSS information.
The quoted 95\% confidence limits are where each curve drops below
the dashed line.
}
\end{figure}

\begin{figure}[tb] 
\centerline{\epsfxsize=8.5cm\epsffile{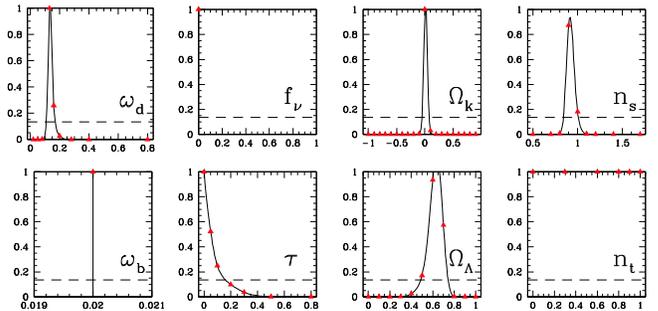}}
\bigskip
\caption{\label{1DnucleoFig}\footnotesize%
Like the previous figure, but adding the 
``concordance'' priors
$\ob=0.02$, $h=0.74\pm 0.08$ and $\fn\sim 10^{-3}$.
}
\end{figure}

Our constraints on individual cosmological parameters are 
listed in Table 1 for three cases.
The best fit model is shown in \fig{CPfig} for case 3.
Constraints are plotted in 
figures~\ref{1DnoFig} and~\ref{1DnucleoFig} for cases 2 and 3.
All tabulated and plotted bounds are 95\% confidence limits\footnote{
Bayesean 95\% confidence limits
are in general those that enclose 95\% of the area.
In this paper, we make the approximation that the boundary of the
confidence region is that where the likelihood has fallen by a factor
$e^{-2}$ from its maximum for 1-dimensional cases (such as 
the numbers in Table 1) and by a factor $e^{-6.18}$ for  
2-dimensional cases 
(such as figures~\ref{1DnoFig} and~\ref{1DnucleoFig}).
As shown in Appendix A of \cite{10par}, this approximation
becomes exact only for the case when the 
likelihood has a multivariate Gaussian form. 
We make this approximation to be consistent with 
the multidimensional marginalization algorithm employed
here (and by most other authors), which is equivalent to the 
integration technique only for the Gaussian case. 
To give the reader a quantitative feeling for the importance of these
issues, we also quote one-sided limis the on $\tau$ and $\nu$
in Table 1, since they have the most asymmetric distributions.
For a detailed discussion of these issues, see \cite {Feldman}.
}.
The first case uses constraints from CMB alone, which are 
still rather weak because of degeneracy problems 
such as the one illustrated 
in \fig{DegenFig}.
The second case combines the CMB information with 
the power spectrum measurements from PSCz, and is seen to
give rather interesting constraints on most parameters
except the tensor tilt $\nt$\footnote{The 
reason that we get no constraints on $\nt$ is that models 
with $\At=0$ fit the data very well, for which varying 
$\nt$ of course has no effect. 
}.
The third case adds three assumptions:
that the latest measurements of
the baryon density $\ob=0.019\pm 0.0024$ 
from Big Bang Nucleosynthesis (BBN) are correct
\cite{Burles98,Burles99},
that the $1\sigma$ constraints on the 
Hubble parameter are $h=0.74\pm 0.08$ 
\cite{Freedman00},
and
that the neutrino contribution is cosmologically negligible.
Since the quoted BBN error bars are much smaller than our $\ob$ grid spacing, 
we simply impose $\ob=0.02$. Also for simplicity, we take the 
errors on $h$ to be Gaussian.
The neutrino assumption is that there is no strong mass-degeneracy between
the relevant neutrino families, and that the Super-Kamiokande atmospheric
neutrino data therefore sets the scale of the 
neutrino density to be $\on\sim\times 10^{-4}-10^{-3}$ 
\cite{Scholberg99}.
We emphasize that this last assumption 
(that the heaviest neutrino
weighs of order the root of the squared mass difference
$\Delta m^2\sim 0.07\,\eV^2$)
is merely motivated by Occam's razor,
not by observational evidence --- 
the best current limits on $\fn$ from other astrophysical
observations (see \cite{Croft99}
and references therein) are
still compatible with $\fn\sim 0.2$.
Rather, we have chosen to highlight the consequences
of this prior since, as discussed below, 
it has interesting effects on other parameters.

For the first 7 parameters listed in Table 1, the 
numbers were computed from the corresponding
1-dimensional likelihood functions (these are plotted 
in \fig{1DnoFig} and \fig{1DnucleoFig} for the 
second and third cases). The best fit value corresponds to the 
peak in the likelihood function and the $95\%$ limits correspond
to where the likelihood function drops below the dashed line 
at $e^{-2}$ of the peak value.
For the remaining parameters listed,
which (except for $b$) are not fundamental parameters in our 11-dimensional grid, 
the numbers were computed as in 
\cite{Jaffe00}
by calculating the likelihood-weighted means and standard 
deviations over the the 
multidimensional parameter space.
Here the tabulated best fit values are this mean and the limits
are the mean $\pm 2\sigma$.

\bigskip
\bigskip
\bigskip

\bigskip
\noindent
{\footnotesize
{\bf Table 1} -- Best fit values and 95\% confidence limits on
cosmological parameters.
The ``concordance'' case combines CMB and PSCz information with
a BBN prior $\ob=0.02$, a Hubble prior $h=0.74\pm 0.08$ and 
a prior that $\fn\sim 10^{-3}$. A dash indicates that no meaningful
constraint was obtained.
The redshift space distortion parameter is
$\beta\equiv f(\Om,\Ol)/b$, where $f$ is the linear growth rate.
$\zion$ is the redshift of reionization, $t_0$ is the present age
of the Universe and $\sum m_\nu$ is the sum of the neutrino masses.
The values labeled as ``best'' are in all cases the ones maximizing the likelihood.
For the numbers below the horizontal line, 
were the limits were computed from
moments as described in the text, the corresponding mean valus are
$h=.53$, $\zion=7$, $t_0=15.6$ (CMB alone),
$b=1.26$, $h=.59$, $\beta=.63$, $\zion=9$, $t_0=13.3$ (CMB+PSCz)
and
$b=1.10$, $h=.68$, $\beta=.51$, $\zion=6$, $t_0=13.4$ (``concordance'').
If the reader wishes to use some of these model pararameters for
other purposes, the numbers to use are thus those in the table, not the ones here
in the caption.
Since the distributions for $\tau$ and $\fn$ are quite asymmetric, we also
quote the 1-sided 95\% limits 
$\tau<0.22$ (CMB only),
$\tau<0.34$, $\fn<0.35$ (CMB+PSCz),
$\tau<0.16$ (``concordance'').
\def\fnp{{$\sim$0}}
\bigskip
{
\begin{tabular}{|l|ccc|ccc|ccc|}
\hline
			&\multicolumn{3}{c|}{CMB alone}
			&\multicolumn{3}{c|}{CMB + PSCz}
			&\multicolumn{3}{c|}{Concordance}\\
Quantity		&Min	&Best	&Max	&Min	&Best	&Max	&Min	&Best	&Max\\
\hline
$\tau$			&0.0	&0.0	&$0.32$	&0.0	&0.0	&$.44$	&0.0	&0.0	&$.16$	\\
$\Ok$			&$-$.69	&$-$.34	&0.05	&$-$.19	&$-$.02	&0.10	&$-$.05	&$-$.00	&0.08	\\
$\Ol$			&$.05$	&.41	&$.92$	&$-$	&.35	&0.76	&.49	&.62	&0.74	\\
$h^2\Od$		&0.0	&.09	&$-$	&.10	&.19	&0.32	&.11	&.13	&0.17	\\
$h^2\Ob$		&.024	&.049	&$.103$	&.020	&.029	&.037	&{\it .02}&{\it .02}&{\it .02}\\
$\fn$			&0.0	&0.0	&$1.0$	&0.0	&.16	&.38	&\fnp	&\fnp	&\fnp	\\
$\ns$			&.91	&1.42	&$-$	&0.86	&.98	&1.16	&0.84	&.91	&1.01	\\
$\nt$			&$-$	&0.0	&$-$	&$-$	&0.0	&$-$	&$-$	&0.0	&$-$	\\
\hline		
$b$			&$-$	&$-$	&$-$	&.75	&1.36	&1.78	&.87	&1.23	&1.33	\\
$h$			&.18	&.39	&.88	&.33	&.57	&.86	&.58	&.63	&.78	\\
$\beta$			&$-$	&$-$	&$-$	&.37	&.59	&.89	&.36	&.47	&.66	\\
$\zion$			&0	&0	&21	&0	&0	&$26$	&0	&0	&20	\\
$t_0$ [Gyr]		&8.4	&18.0	&23.0	&9.6	&13.1	&17.0	&12.1	&14.0	&14.6	\\
$\sum m_\nu$ [eV]	&0	&0	&17	&0	&2.7	&7.6	&\fnp	&\fnp	&\fnp	\\
\hline		
\end{tabular}
}
}

\subsection{Effects of priors}

Case 3 in Table 1 is but one from a selection of about 20
different priors that we tried experimentally. The reason that we have 
chosen to highlight this one is that 
the nucleosynthesis constraint was the one that had the greatest impact
on the results. Otherwise, the 
joint CMB+PSZc constraints were remarkably robust to prior assumption.
Imposing priors such as flatness ($\Ok=0$), 
no tensors ($r=0$), no tilt ($\ns=1$), no reionization ($\tau=0$),
and a reasonable Hubble parameter
(we tried both $h=0.74\pm 8$ at 65\% 
and the weaker constraint $50<h<100$ at 95\%),
both alone and in various combinations, 
has little effect.
The fact that the best fit parameter
values are not appreciably altered reflects that these priors
all agree well with what is already borne out by the
CMB+PSCz data: $\Ok\sim r\sim\tau\sim 0$, 
and $\ns=1$.
The fact that these priors do not shrink the error bars
much on other parameters indicates that the
PSCz has already broken the main CMB degeneracies.

\begin{figure}[tb] 
\centerline{\epsfxsize=8.5cm\epsffile{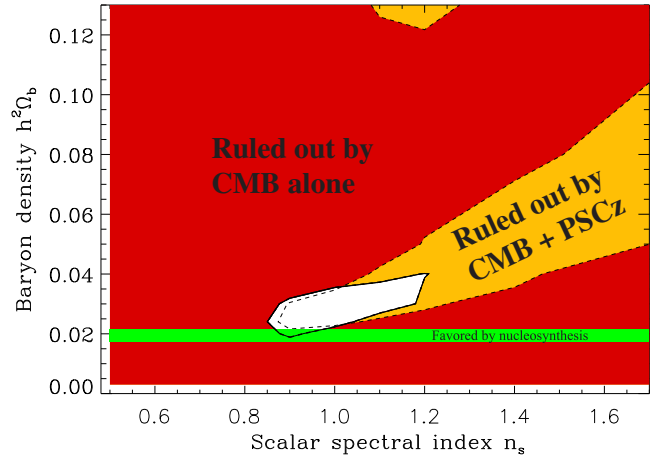}}
\bigskip
\caption{\label{nsobFig}\footnotesize%
Constraints in the $(\ns,\ob)$-plane. Note that PSCz not only shrinks
the allowed region (white), but also pushes it slightly down to the left
(the dashed line indicates the CMB-only boundary).
}
\end{figure}

\begin{figure}[tb] 
\centerline{\epsfxsize=8.5cm\epsffile{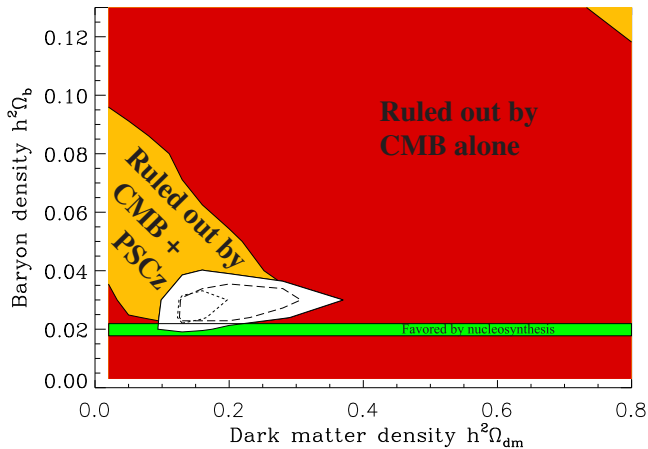}}
\bigskip
\caption{\label{odobFig}\footnotesize%
Constraints in the $(\od,\ob)$-plane.
As in the previous figure, adding PSCz prohibits 
high baryon solutions and allows slightly 
lower $\ob$-values than CMB alone.
The dashed curve within the allowed (white) region
show the sharper constraint obtained when 
imposing the priors
for a flat, scalar scale-invariant model 
($\Ok=r=0$, $\ns=1$). 
The dotted curve shows the effect of requiring 
negligible neutrino density ($\fn\sim 0$) in addition.
}
\end{figure}

The nucleosynthesis prior has a greater influence
because it does not agree all that well with what the
CMB+LSS data prefer. Although adding PSCz is seen 
to pull down the preferred baryon density slightly, 
reducing the 95\% lower limit from 0.024 (CMB only) to
0.020 (CMB+PSCz), the preferred value of 0.028 still
exceeds the BBN value.
It is well-known that CMB likes either a high baryon density
or a red-tilt $n<1$ because of the low second acoustic peak
\cite{Lange00,boompa,Bambi00,observables,Hu00,White00},
and the CMB exclusion region to the lower right in 
\fig{nsobFig} illustrates this tradeoff.
Enforcing the BBN baryon value therefore shifts the preferred
tilt-range away from the scale-invariant $\ns\sim 1$ case to
$\ns\sim 0.9$. Since $\ob$ is one of the
few parameters affecting the relative heights of the
acoustic peaks (together with $\od$, $\ns$ and, marginally, $\fn$), 
eliminating the uncertainty in $\ob$ with the
BBN prior also tightens the constraints on these other parameters.
In particular, the link between $\ob$ and $\od$ is illustrated in 
\fig{odobFig}.

\begin{figure}[tb] 
\centerline{\epsfxsize=8.5cm\epsffile{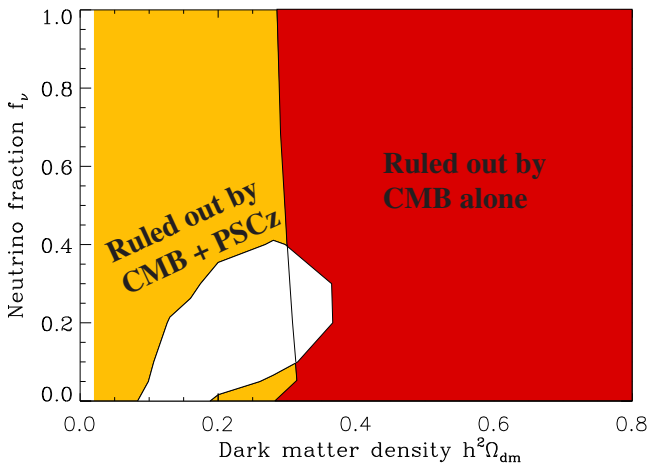}}
\bigskip
\caption{\label{odfnFig}\footnotesize%
Constraints in the $(\od,\fn)$-plane.
The shape of the allowed (white) region explains why the prior 
$\fn=0$ tightens the upper limit on the dark matter density.
The vertical line shows the CMB-only boundary before PSCz is added.
}
\end{figure}

We found one additional prior that had a non-negligible effect:
that on neutrinos. As illustrated in \fig{odfnFig}, inclusion of 
neutrinos substantially weakens the upper limits on the dark matter
density.
Since the neutrino fraction $\fn$ has only a weak effect on the CMB, 
this effect clearly comes from LSS. 
A larger dark matter density $\od$ pushes matter-radiation
equality back to an earlier time, shifting the corresponding 
turnover in $P(k)$ to the right and thereby increasing the
ratio of small-scale to large-scale power.
Increasing the neutrino fraction counteracts this by suppressing
the small-scale power (without affecting the CMB much), 
thereby weakening the upper limit on $\od$.
Imposing the prior $\fn=0$ alone,
without nucleosynthesis or Hubble priors,
tightens the CMB+PSCz constraint $\od<0.32$ from Table 1 to $\od<0.19$.

\begin{figure}[tb] 
\centerline{\epsfxsize=8.5cm\epsffile{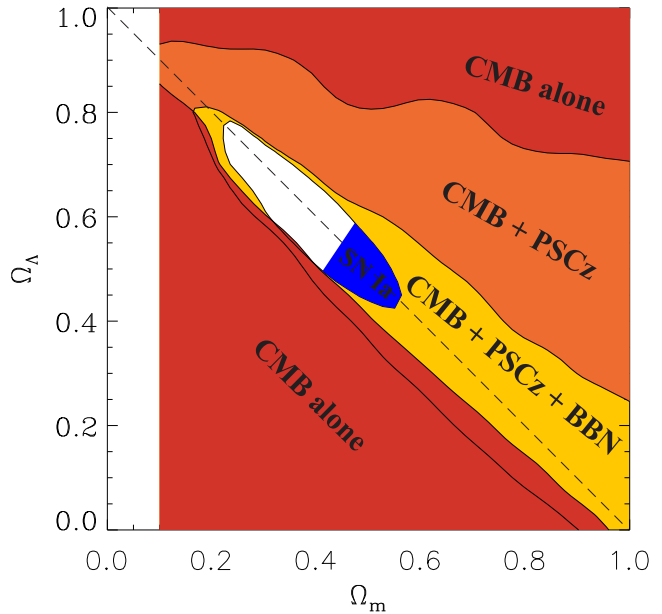}}
\bigskip
\caption{\label{OmOlFig}\footnotesize%
Constraints in the $(\Om,\Ol)$-plane.
The shaded regions are ruled out at 95\% confidence by the
information indicated. The allowed (white) region is seen to be centered
around flat models, which fall on the dashed line.
}
\end{figure}

Since the constraints on $\od$ are tightened by fixing
both $\ob$ (\fig{odobFig}) and $\fn$ 
(\fig{odfnFig}), the ``concordance'' case in Table 1 
gives quite tight constraints on the dark matter density.
The $h$ prior helps turn this $\od$ constraint into a measurement 
of $\Od$, and the measurement $\Ok\sim 0$ therefore gives
an indirect constraint on the cosmological constant via
$\Ol=1-\Ok-\Od-\Ob$. This is illustrated in \fig{OmOlFig},
where the concordance constraints close off the allowed region by placing
a lower limit on $\Ol$. This lower limit on $\Ol$ goes away if we
drop either the $\ob$-prior or the $\fn$-prior.

\subsection{Constraints on other parameters}

\subsubsection{Hubble parameter}

Table 1 shows that the constraints on the Hubble parameter $h$ are
quite weak for the CMB+LSS case. However, the bound 
$h<0.78$ (95\%) for the concordance case is noteworthy since it is 
much stronger than (and hence independent of) the 
prior $h=0.74\pm 0.08$ that was used.

\subsubsection{Ionization redshift}

In the approximation that the diffuse hydrogen in the
Universe became fully ionized rather
abruptly at a redshift $\zion$, 
this quantity is well approximated by
(\eg, Peebles 1993)
\beq{zionEq}
\zion = 8.9\>\left({\tau h\over\ob}\right)^{2/3}\Om^{1/3}
\eeq
as long as $\zion\gg 1$.
Although the constraints
in Table 1
agree well with what is predicted in
recent simulations, more extreme models are seen to be ruled out.
Earlier CMB constraints on $\zion$ were studied in 
\cite{Liddletau,Venkatesan}.

\subsubsection{Age of Universe}

The ability to place constraints in the $(\Om,\Ol)$-plane
is allowing the age of the Universe to be predicted with
improved accuracy 
\cite{Lineweaver99,Jaffe00}.
The 95\% confidence interval for our concordance case, 
12.1--14.6 Gigayears, is consistent with direct age determinations
from, \eg, globular clusters \cite{Sarajedini97,Rosenberg99}.

\section{DISCUSSION}

\label{ConclusionsSec}

We have presented a method for rapid and accurate power spectrum calculation 
for large numbers of CMB models
and used it to constrain jointly 11 cosmological parameters from current 
CMB and galaxy clustering data. 
Perhaps the most interesting results of this paper are the numbers themselves,
listed in the CMB+LSS columns of Table 1, and their striking robustness to 
imposing various priors. A superficial glance at the constraint figures 
might suggest that little has changed since the first analysis  
of Boomerang + Maxima \cite{boompa}, or even since the pre-Boomerang analysis
of \cite{10par}, since the plots look rather similar. 
However, whereas these earlier
papers obtained strong constraints only with various poorly 
justified priors such as no tensors, no tilt or no curvature, 
the joint CMB + LSS data are now powerful enough to 
speak for themselves, without needing any such prior props.

\subsection{New public software}

Our new power spectrum calculation 
method accelerates CMBfast by about a factor $10^3$. 
It is accurate
to about 1--2\% for $C_\l$ on all angular scales and to about 1\% for $P(k)$ on 
scales $k<0.15\,h\,\Mpc^{-1}$. Since this is roughly the 
intrinsic accuracy level of CMBfast 
itself, there is no reason not to take advantage of this technique 
when constraining cosmological parameters.
A modified version of CMBfast incorporating our $k$-split method will be
made publicly available at 
www.sns.ias.edu/$\sim$matiasz/CMBFAST/cmbfast.html.

\subsection{Caveats}
\label{kcutSec}

Let us now discuss assumptions and approximations that underly our analysis.

For both the CMB and LSS likelihood calculations, the percent
level inaccuracies in our power spectrum computation are likely to have
a negligible effect.
Indeed, the least accurate models tend to be wild ones 
that are inconsistent with the data in any case.
On the LSS side, the dominant uncertainties are likely to be
related to the measured $P(k)$ instead.
Specifically, our use of the 
measurements of
the PSCz real-space power \cite{pscz} assumed both that linear perturbation
theory was valid and that the bias was scale-independent on these scales.
Let us now discuss both of these assumptions in turn.

To assess the possibility that nonlinear effects at
$k\sim 0.1$--$0.3 \, h\,\Mpc^{-1}$ had tainted our results, 
we repeated the entire analysis
twice, discarding all $P(k)$-measurements for $k$
exceeding $0.2\,h\,\Mpc^{-1}$ and $0.1\,h\,\Mpc^{-1}$, respectively.
The upper limit $\fn<0.38$ on the neutrino fraction
was weakened for the 0.2 case and 
went away completely for the 0.1 case.
Thus the upper limit on neutrinos is sensitive to information
at mildly nonlinear scales.
The upper limit on $\ns$ 
was also weakened as the lever arm shortened, 
but only very slightly, from $\ns<1.16$ at $0.3 \, h \, \Mpc^{-1}$ to
$\ns<1.18$ and $\ns<1.19$ at $0.2$ and $0.1 \, h \, \Mpc^{-1}$, respectively.
Other constraints were
less affected.

The redshift distortion study reported
in \cite{pscz}
suggested that,
while nonlinear effects are visible in the galaxy-velocity power spectrum
already at scales $k \approx 0.15 \, h \, \Mpc^{-1}$,
linear theory is probably a fair approximation down to
$k \approx 0.3 \, h \, \Mpc^{-1}$.
A subsequent study \cite{pscz2} of the nonlinear power spectrum of PSCz
has shown that the galaxy power spectrum is likely to be antibiased
relative to the matter power spectrum at translinear scales,
a conclusion previously arrived at by \cite{Kravtsov99,Jenkins98}.
Such antibias tends to cancel the effects of nonlinearity in the
matter power spectrum, making the galaxy power spectrum appear
similar to the linear matter power spectrum down to 
$k \approx 0.3 \, h \, \Mpc^{-1}$.
The fact that the relative bias between APM and PSCz galaxies
is consistent with being constant,
$b_{\rm APM}/b_{\rm PSCz} \approx 1.15$,
for $k \le 0.3 \, h \, \Mpc^{-1}$
\cite{Peacock97,pscz2}
adds further circumstantial evidence suggesting that
the galaxy power spectra are near linear at these scales.

We therefore conclude that including the PSCz data at
$k = 0.1$--$0.3 \,h\,\Mpc^{-1}$
does not bias the results significantly,
and that, aside from the constraint on neutrinos,
the information from PSCz is not dominated by these last few bins.

Although most theoretical work has suggested that the bias $b$ is unlikely 
to vary much on linear scales, we must still be open to the possibility
of scale-dependent bias masquerading as a 
cosmological effect. For instance, one might imagine that 
more luminous galaxies are more highly biased, and that
they carry a greater statistical weight for the leftmost $k$-bands since
they remain in the magnitude-limited 0.6 Jy sample even at great distances.
Such a luminosity bias could masquerade as a slight red-tilt $\ns<1$.
However,
three recent studies
\cite{Szapudi00,Beisbart00,Hawkins00}
all conclude that PSCz galaxies of different luminosity cluster similarly.

On the CMB side, a long list of approximations in our treatment were discussed
in \cite{10par}, involving both the likelihood 
calculation \cite{BJK98}
and the marginalization.
The dominant limitation is likely to be that we have not included
the full window functions and slight 
band-band correlations of 
Boomerang and Maxima, but this is unfortunately not possible until the
relevant Fisher matrices are made public by the two experimental teams.

Positive correlations between neighboring data points makes is easier
(in terms of chi-squared cost) to shift the overall height of say 
the first peak up and down, thereby weakening the constraints
on parameters that are measured mainly from peak heights 
($\ob$, $\ocdm$, $\ns$). This effect is similar to that of calibration errors,
which we did include.
The fact that our best fit models tend to predict a lower
first peak than much of the data illustrates the effect of calibration
errors and also shows that an extremely high first peak is hard to achieve
given the other constraints.

The extraction of constraints on individual parameters
from the multidimensional likelihood function can be
done in a number of different ways.
Three ways of marginalizing are discussed in \cite{10par}
(by integration, by maximization over the grid 
and by maximization over a smooth interpolating function),
all of which have been used in the recent literature.
An encouraging indication that
our results are insensitive to the method of marginalization
comes from comparing the marginalized constraints from
Table 1 with those obtained without any marginalization, 
from the above-mentioned computation of means and standard 
deviations by summing over the grid.
These completely different methods give quite similar results for
all well-constrained parameters. A typical example is the $2\sigma$
range for $\od$ for the ``concordance'' case, coming out as 
$0.110<\od<0.170$ (marginalized) and 
$0.105<\od<0.172$ (from moments).
Even for $\tau$, the parameter with the most non-Gaussian likelihood,
the $2\sigma$ ``concordance''
upper limits are similar: 0.161 and 0.163, respectively.

Finally, although we repeatedly referred to the ``no prior'' case for our
CMB + LSS analysis, it is important to bear in mind that there
is strictly speaking no such thing as no priors.
Specifically, all our calculations assumed that the adiabatic
inflationary paradigm is correct. We also assumed that the dark energy
was a cosmological constant rather than some form of ``quintessence''
with a different equation of state. Finally, the edges of our parameter
grid imposed a hard-wired top hat prior. This had a negligible effect on our
results for all parameters except one, since the likelihood dropped to
negligible values well before reaching the boundary of our 11-dimensional 
parameter space. The one exception involved $\Ol$, since 
\fig{OmOlFig} illustrates that negative 
$\Ol$ cannot be excluded except when either 
BBN or SN 1a information is included.

\subsection{Comparison with other recent work}

Our results agree fairly well with the recent constraints from other 
groups
\cite{Lange00,Bambi00,observables,Jaffe00,Kinney00}.
The analysis most comparable to ours is that of 
Jaffe {\etal} \cite{Jaffe00}.
That study
limits LSS-information to that incorporated in a
shape parameter and
a normalization parameter, and
uses a smaller
CMB data set, limiting the analysis to COBE, Boomerang and Maxima.
The main effect of this culling is likely to enter on
scales $50\simlt\l\simlt 200$, covering the rise toward the first
acoustic peaks, where Boomerang and Maxima are both sample 
variance limited and other experiments have covered a substantially
larger sky area. 
In addition, \cite{Jaffe00}
limit their parameter space to no neutrinos and no tensors
($\fn=r=0$), employ a different 
numerical marginalization scheme
and include the above-mentioned proprietary band correlations.
The fact that our results agree so well despite all these technical differences
is quite reassuring, indicating that the data are 
now good enough to make the results insensitive to details of method.
The stronger upper CMB+LSS limit on $\od$ obtained in \cite{Jaffe00},
which indirectly gives $\Ol\simgt 0.5$ as described above, 
is presumably due to 
their no-neutrino prior $\fn=0$, as can be understood from \fig{odfnFig}.
The fact that their CMB-only constraints are stronger than ours
traces back to their no-tensor $r=0$ prior --- this prior eliminates
the $\ns-r-\ob$ degeneracy that we needed the full PSCz data to break,
and automatically excludes models such as the one shown in \fig{DegenFig}.

Our results also agree well with those from the likelihood analysis of 
Kinney {\etal} \cite{Kinney00}. Although this analysis has $\fn=0$, a prior for $\nt$ 
and a limited $\tau$-range, it includes a thorough treatment of tensor
modes and maps out the $(n_s,r)$-plane in detail. The study 
finds that quite blue-tilted models
are allowed when $r$ is large, precisely the effect that degrades our
CMB-only constraints --- see 
also \cite{Melchiorri99,9par}.

\subsection{Towards a refined concordance model}

It is well-known that different types of measurements can
complement each other by breaking degeneracies. 
However, even more importantly, multiple data sets allow numerous
consistency checks to be made. The present results allow a number of
such tests.

\subsubsection{Baryons}

Perhaps the most obvious one involves the baryon fraction. 
Although there is still some tension between BBN (preferring 
$\ob\sim 0.02$ and CMB+LSS (preferring $\ob\sim 0.03$),
an issue which will undoubtedly be clarified by improved data within 
a year\footnote{
A number of possible theoretical explanations 
for the slight mismatch have been discussed in the recent literature
\cite{Lesgourgues00,Hannestad00,Hansen00,Orito00,Esposito00a,Esposito00b,Esposito00c,Kaplinghat00}.
Another possibility is clearly that the favored value from CMB 
will shift as data improve.
Recent measurements of a high deuterium abundance in the interstellar 
medium make it unlikely that 
the standard BBN value will creep above $\ob=0.025$ \cite{Jenkins99,Sonneborn00},
and a new QSO deuterium absorption study 
reproduces $\sim 0.02$ \cite{Tytler00}.
},
the most striking point is that the methods 
agree as well as they do. That one 
method involving nuclear physics when the Universe was 
minutes
old and another involving plasma physics more than 
100{,}000 years
later give roughly consistent answers, despite involving
completely different systematics, can hardly be described as anything short
of a triumph for the Big Bang model. 
 
It is noteworthy that our addition of LSS 
information pulls down the baryon value slightly, so that
a BBN-compatible value $\ob=0.02$ is now within the 95\% confidence interval.
Part of the reason that that the CMB alone gave a stronger lower 
limit may be a reflection
of the Bayesian likelihood procedure employed in this and all other recent papers
on the topic: when a large space of high $\ob$-values are allowed, the relative
likelihood for lower values drops.

In all three cases listed in Table 1, the best fit model is consistent
with the data in the sense of giving an acceptable $\chi^2$-value.
The ``CMB only'' case gives $\chi^2\approx 70.3$ for 87 degrees of 
freedom. The ``CMB+PSCz'' case gives $\chi^2\approx 88.7$, which rises to
95.6 for the ``concordance'' case
--- all for
$\approx 109 - 11$
degrees of freedom.
The effective number of degrees of freedom might be a few larger than this,
since some of the 11 parameters had little effect, but
even taking this into account, all fits are good in the sense of 
giving reduced $\chi^2$-values of order unity.

Apart from BBN, our baryon value also agrees with the range
$0.007\simlt\Ob\simlt0.041$
inferred from a low-redshift inventory \cite{Fukugita98}
and the range
$0.015\simlt\ob\simlt 0.03$
at redshifts of a few from the Ly$\alpha$ forest
\cite{Rauch00,Weinberg97,Zhang98,Theuns99}.
The inferred baryon fraction $\ob/\od\sim 15\%$
agrees well with that inferred from 
galaxy clusters
\cite{Mohr00,Durrer00} 
for reasonable $h$-values.

It is difficult to contemplate the PSCz data in 
\fig{PSCzFig} without wondering whether they show
evidence for baryonic wiggles in the matter power spectrum. 
Intriguingly,
the location and amplitude of the wiggles
in the PSCz power spectrum
fit well to a flat, pure baryon model
with $\Ol = .86$ and $\Om = \Ob = 0.14$ (for $h = 0.7$),
albeit with a large blue tilt, $n_s = 2$,
but agrees poorly with the CMB.
The PSCz data are also entirely consistent
with a wiggle-free spectrum. 

\subsubsection{Dark energy}

Another important cross-check involves the cosmological constant.
Although the constraint $0.49<\Ol<0.74$ from Table 1 does not involve 
any supernova information, it agrees nicely with the recent
accelerating universe predictions from 
SN 1a \cite{Perlmutter98,Riess98}. This agreement is illustrated in 
\fig{OmOlFig}, which shows the SN 1a constraints from 
\cite{WhiteConcordance} combining the data from 
both teams.
As frequently pointed out, the conclusion $\Om\sim 0.35$ also
agrees well with a number of other observations, \eg, 
the cluster abundance at various redshifts 
\cite{Bahcall98,Eke98,Henry00,Mohr00}
and cosmic velocity fields 
\cite{Zehavi99},
although there is still some internal controversy in these
two areas (see, \eg, \cite{Liddle99}).

\begin{figure}[tb] 
\centerline{\epsfxsize=8.5cm\epsffile{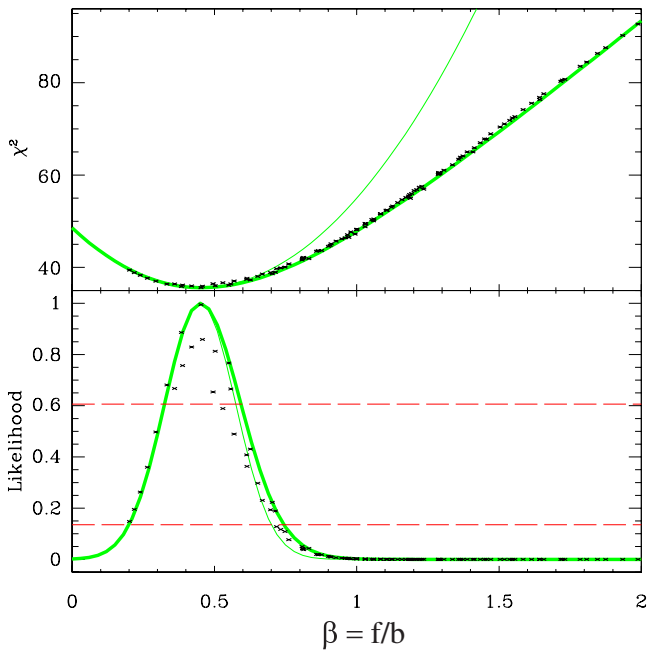}}
\bigskip
\caption{\label{betaFig}\footnotesize%
The upper panel shows $\chi^2$ for 
the linear redshift distortion parameter beta
measured from the PSCz data.
Specifically, $\chi^2$ was computed using the 
galaxy-galaxy, galaxy-velocity and velocity-velocity
power spectra but no CMB information, then 
minimized over all parameters except
$\Ok$ and $\Ol$. The fact that the resulting 
$\chi^2(\Ok,\Ol)$ falls near a 1-dimensional curve with little scatter 
shows that it essentially
only depends on one particular function of these
two parameters, the linear growth rate $f(\Om,\Ol)$
(equivalently the redshift distortion parameter $\beta=f/b$).
The best fit curve (thick line) has a slower rise for high $\beta$ than
a parabola (thin line).
The lower panel shows the corresponding likelihood 
$\L\propto e^{-\chi^2/2}$,
and the $1\sigma$ and $2\sigma$ confidence limits are where this
(thick) curve crosses the dashed lines.
}
\end{figure}

\subsubsection{Bias}

A third cross-check is more subtle but equally striking,
involving the bias of the PSCz galaxies --- we can measure
it in two completely independent ways.
One is by comparing the amplitude of the 
CMB and galaxy power spectra, which gives the constraints listed
in Table 1.
The other way is via the linear redshift space distortion 
parameter $\beta=f(\Om,\Ol)/b$, where
$f(\Om,\Ol) \approx\Om^{0.6}$ is the dimensionless linear growth rate
\cite{Peebles80,Lahav91,Hamilton00}.
We therefore remove the CMB data (which eliminates the 
$\beta$-constraints from Table 1) and add in their place
the two PSCz power spectra from \cite{pscz} that we discarded 
above. These are the galaxy-velocity and velocity-velocity
power spectra from the stochastic bias formalism, roughly speaking
corresponding to the quadrupole and hexadecapole of the 
redshift space distortions.
The redshift distortion parameter for the PSCz galaxies was measured
to be $\beta=0.41^{+.13}_{-.12}$ \cite{pscz},
a result in good agreement with the PSCz team's own most recent measurement,
$\beta=0.39 \pm 0.12$
\cite{Taylor00}.
Both of these measurements involved a limited marginalization
over the power spectrum. Here we marginalize the full likelihood
function over all our cosmological parameters except $\beta$.
As can be seen in \fig{betaFig}, this gives the $1\sigma$ 
measurement 
$\beta=0.45^{+.14}_{-.12}$.
The fact that this agrees so well with the corresponding 
$1\sigma$ measurement $\beta=0.51\pm 0.08$ from Table 1 for the
concordance case means that a highly non-trivial consistency test
has been passed.

\subsection{Concordance}

In conclusion, the simple ``concordance'' model 
in the last columns of Table 1 (plotted in \fig{CPfig})
is at least marginally consistent with all basic cosmological constraints, 
including CMB, PSCz and nucleosynthesis.
Specifically, as discussed above, our calculations show that 
it 
has passed three non-trivial consistency tests.
Moreover our concordance model is encouragingly robust towards
imposing a score of prior constraints in various combinations.
Cosmology seems to be on the right track!

\bigskip
The authors wish to thank 
Ang\'elica de Oliveira-Costa, Daniel Fisher, Brad Gibson, Wayne Hu,
William Kinney, Arthur Kosowsky, and Nikhil Padmanabhan
for useful discussions and helpful comments.
Support for this work was provided by
NSF grant AST00-71213, 
NASA grants NAG5-7128 
and NAG5-9194,
the University of Pennsylvania Research Foundation,
and 
Hubble Fellowship HF-01116.01-98A from 
STScI, operated by AURA, Inc. 
under NASA contract NAS5-26555. 

\appendix

\section{BASIC METHOD IMPROVEMENTS}

\label{AppendixMethodSec}

As mentioned, our method is based on the one described in \cite{10par},
but with a number of extensions and improvements as detailed below.
It consists of the following steps:
\begin{itemize}
\item Compute power spectra $C_\l$ and $P(k)$ for a grid
of models in our 11-dimensional parameter space.
\item Compute a likelihood for each model that quantifies how well it fits the
data.
\item Perform 11-dimensional interpolation and marginalize to obtain
constraints on individual parameters and parameter pairs.
\end{itemize}
Our main improvement over \cite{10par} is in step 1.
We improve the technique for accelerated CMB power
spectrum calculation so that it becomes essentially as 
accurate as CMBfast itself, simply about $10^3$ times faster.
We also add a simple but accurate technique to compute the grid 
of matter power spectra rapidly.

\subsection{Improved choice of parameters}

Reference \cite{10par} explored the 10-dimensional parameter space involving
the reionization optical depth $\tau$, 
the primordial amplitudes $\As$, $\At$ and tilts $\ns$, $\nt$ 
of scalar and tensor fluctuations, and 
various contributions $\Omega_i$ to critical density.
The $\Omega_i$ included were for 
curvature $\Ok$, vacuum energy $\Ol$, cold dark matter $\Oc$, 
hot dark matter (neutrinos) $\On$ and baryons $\Ob$.
Since it is computationally advantageous to work with 
parameters that are closely linked to the most important physical
processes involved, reference \cite{10par} used the parameter vector
\beq{OldpEq}
\p\equiv(\tau,\Ok,\Ol,\ocdm,\ob,\on,\ns,\nt,\As,\At),
\eeq
where the physical densities 
$\omega_i\equiv h^2\Omega_i$.

Unless $\On\ll 1$, the neutrinos left over from the early Universe were
heavy enough to be 
fairly non-relativistic during the processes that created the
acoustic peaks, and thereby had almost the same effect as cold dark matter
on the CMB. The CMB power spectrum therefore depends mainly on 
the total
nonbaryonic
(cold$+$hot) dark matter density $\od\equiv\ocdm+\on$ 
and only quite weakly
on the hot fraction $\fn\equiv\on/\od$. We therefore 
replace our old parameters $(\ocdm,\on)$ by $(\od,\fn)$.
This allows us to accurately compute the weak $\fn$-dependence of the 
scalar CMB power spectrum by using a coarse grid 
$\fn=$0.0, 0.3, 1.0 and interpolating.
Moreover, the tensor fluctuations are essentially independent of $\fn$,
so we only compute them for $\fn=0$.

In this paper, we need to add one more parameter, 
relating the theoretically predicted  
power spectrum of matter $P(k)$ to
that of PSCz galaxies $P_g(k)$ on large scales. 
This parameter is the bias $b\equiv [P_g(k)/P(k)]^{1/2}$
from the stochastic bias formalism
\cite{Pen98,Dekel99,bias}.
Although it can in principle depend
on scale, we will assume that it is constant on the large scales 
that we consider
\cite{Coles93,Fry93,Scherrer98,Coles99,Kravtsov99,Colin99,Narayanan00,Benson00}.
We will make no assumptions about the
value of $b$, however, and therefore marginalize over this parameter
before quoting constraints on the other ten.
In summary, we use the parameter vector 
\beq{pEq2}
\p\equiv(\tau,\Ok,\Ol,\od,\ob,\fn,\ns,\nt,\As,\At,b).
\eeq
Note that the Hubble constant is not a twelfth independent
parameter, since 
\beq{hEq}
h = \sqrt{\od+\ob\over 1-\Ok-\Ol}.
\eeq
We wish to probe a large enough region of parameter space to cover
even quite unconventional models. This way, constraints from non-CMB
observations can be optionally included by explicitly multiplying 
the likelihood function $\L(\p)$ by a Bayesian prior rather than being 
hard-wired in from the outset.
To avoid dealing with prohibitively many models,
we use a roughly logarithmic
grid spacing for $\om$, $\ob$ and $\on$, 
a linear grid spacing for $\Ok$ and $\Ol$,
a hybrid for $\tau$, $\fn$, $\ns$ and $\nt$, 
and (as described below) a continuous grid for $\As$, $\At$ and b.

The recent progress in CMB accuracy has been so dramatic
that the grids used in some recent papers \cite{boompa,Bambi00}
are already 
almost too sparse to accurately sample the small
allowed regions of parameter space. We therefore modify the
grid from \cite{boompa} to zoom in on the favored parameter ranges while still retaining 
some outlier points to be on the safe side.
We let the parameters take on the following values:
\begin{itemize}
\item $\tau=0, 0.05, 0.1, 0.2, 0.3, 0.5, 0.8$ 
\item $\Ol=0, 0.1, ...., 1.0$ 
\item $\Ok$ such that $\Om\equiv 1-\Ok-\Ol = 0.1, 0.2, ..., 1.0$
\item $\od=.02, .05, .08, .13, .16, .20, .28, .40, .80$ 
\item $\ob=.003, .013, .016, .020, .024, .03, .04, .05, .08, .13$ 
\item $\fn=0, 0.05, 0.1, 0.2, 0.3, 0.4, 0.6, 0.8, 1.0$ 
\item $\ns=0.5, 0.7, 0.8, 0.9, 1.0, 1.1, 1.2, 1.4, 1.7$ 
\item $\nt=-1.00, -0.70, -0.40, -0.20, -0.10, 0$ 
\item $\As$ is not discretized
\item $\At$ is not discretized
\item $b$ is not discretized
\end{itemize}
Note that the extent of the $\Ok$-grid depends on $\Ol$, giving 
a total of $10\times 11=110$ points in the $(\Om,\Ol)$-plane.
Our discrete grid thus contains 
$7\times 110\times 9\times 10\times 9\times 9\times 6=33,679,800$ models.
As in \cite{10par}, the main limitation on 
this grid size is disk space rather than CPU time.

\subsection{The three basic spectra and their normalization}

It is convenient to write the two power spectra $\Cl$ and $P_g(k)$ 
that we can measure as
\beqa{CNormalizationEq}
\Cl	&=& \As\Cl^{\rm scalar} + \At\Cl^{\rm tensor},\\\label{PNormalizationEq}
P_g(k)	&=& \As b^2 P(k),
\eeqa
where the three basic power spectra 
$\Cl^{\rm scalar}$, $\Cl^{\rm tensor}$ and $P(k)$
are all normalized consistently, corresponding 
to a fixed amplitude of the gravitational potential $\psi$ when each
mode is outside horizon. The default output of CMBfast first
normalizes the CMB output to COBE, 
but the new version allows output of
the raw unnormalized spectra that we need.  
$\Cl^{\rm scalar}$ depends on $(\tau,\Ok,\Ol,\od,\ob,\fn,\ns)$,
$\Cl^{\rm tensor}$ depends on $(\tau,\Ok,\Ol,\od,\ob,\nt)$
and 
$P(k)$ depends on $(\Ok,\Ol,\od,\ob,\fn,\ns)$,
so we need to compute three separate grids of models of dimensionality
7, 6 and 6, respectively.
We describe a fast and accurate way of doing this in 
Appendices B and C.

\subsection{Likelihoods and marginalization}
\label{LikelihoodSec}

We compute the CMB likelihood exactly as in \cite{10par}, \ie,
using the first results from
Maxima and Boomerang as well as all prior experiments 
\cite{Gawiser00}
(shown in
\fig{DegenFig})
and
taking into account the effect of calibration errors.

For the PSCz galaxy power spectrum, we use only the band power
measurements plotted in black in \fig{PSCzFig}, omitting the
ones further to the right. 
This is a subset of the measurements from
\cite{pscz}
that includes information only at scales
$k<0.3\,h\,\Mpc^{-1}$ to ensure that we stay clear of nonlinear effects. 
As a further precaution, we examine how the results change when 
this cut is further sharpened in \sec{kcutSec}.
We approximate the corresponding likelihood function by the 
multivariate Gaussian $\L_{\rm lss}\propto e^{-{1\over 2}\chi^2}$,
where $\chi^2$ is computed using the measurements in 
\fig{PSCzFig}.
Each point in \fig{PSCzFig} represents an uncorrelated measurement
of the power in a well-defined band
whose FWHM is indicated by the horizontal bar.
In computing the likelihood,
we take into account the detailed form of each band-power window.

The joint likelihood function is obtained by multiplying the
CMB and LSS likelihoods.
Throughout this paper we marginalize over
the amplitudes $\As$, $\At$, and the bias factor $b$.
Equations\eqn{CNormalizationEq} and\eqn{PNormalizationEq}
show that we can equivalently marginalize
over these parameters separately for the CMB and LSS likelihoods
before multiplying them together, which simplifies the calculations in
practice.

\section{METHOD FOR COMPUTING $C_\l$}

We compute $\Cl^{\rm tensor}$ as described in \cite{10par}:
by running CMBfast merely for a coarser grid in $\od$ and $\ob$
(on which the dependence is weak) and interpolating onto 
our full grid using a regularized multidimensional spline.
Since $\Cl^{\rm tensor}$ only contributes to the first few hundred multipoles, 
it is much faster to compute than $\Cl^{\rm scalar}$, which is the real challenge.

\subsection{The $k$-space split}

The idea introduced in \cite{9par} and \cite{10par} was roughly speaking
to compute the $\l\simlt 100$ and $\l\simgt 100$ parts of
$\Cl^{\rm scalar}$ separately and splice them together afterwards.
The former can be computed just as fast as $\Cl^{\rm tensor}$,
since its dependence on $\od$ and $\ob$ is weak and it is 
essentially independent of $\fn$.
The latter can be computed rapidly as well, since
the only effect of $\Ok$ and $\Ol$ is to shift
it sideways in a known way. 
To a decent approximation, the $\tau$-dependence can be incorporated
analytically as well, as simply a multiplication by $e^{-2\tau}$.
However, since there is a small bump of regenerated power
from the new last scattering surface (which moves
to larger $\l$ as $\tau$ is increased), we opt to include
$\tau$ explicitly this time --- as mentioned, the algorithm is so
fast that we are limited by disk space rather than CPU time anyway.
In short, the high-$\l$ part of $\Cl^{\rm scalar}$ only needs to be
computed on a 5-dimensional grid spanned by
$(\tau,\od,\ob,\fn,\ns)$. Moreover, this is really only 
4 ``hard'' parameters, since CMBfast treats multiple 
$\ns$-values simultaneously with no slowdown.

Although this approximation works well, 
it is typically only accurate to 5--10\% or so.
The main problem is with the splicing itself, since 
projection effects alias power from a given physical scale
to quite a broad range of $\l$-values, blurring the 
separation between the low and high grids.
The new method that we present here bypasses this
problem by making the split directly in $k$-space, where
the actual physics takes place.
Specifically, we modify CMBfast to save only the contribution 
to $\Cl$ from below or above a certain wave number 
$k_*$ when it integrates the Boltzmann equation.

The ``low'' $(k<k_*)$ contribution corresponds
to fluctuations on scales outside the
horizon at recombination. This makes it almost independent of 
the causal microphysics that creates the familiar acoustic peaks, 
\ie, almost independent of $\od$, $\ob$ and $\fn$. 
Rather, it is dominated by what happens at low redshift
(the late integrated Sachs-Wolfe effect, reionization, \etc).

In contrast, the ``high'' $(k>k_*)$ contribution is essentially
unaffected by low redshift effects, so $\Ok$ and $\Ol$ 
(or some other dark energy component that was negligible 
at $z\simgt 10^3$) will merely take the pattern put in place
at $z\sim 10^3$ and shift it sideways according
to the angle-distance relationship. 
Suppressing the other 9 parameters, we thus have
\beq{ShiftEq}
\Cl^{\rm scalar}(\Ok,\Ol) 
\approx \Clo(\Ok,\Ol) + C_{\l'}^{\rm high}(0,0),
\eeq
where $\l'\equiv[\dA(\Ok,\Ol)/\dA(0,0)]\l$ and 
$\dA$ is the angular diameter distance to the last scattering
surface in $\Mpc$ (not in $h^{-1}\Mpc$).

We tune the choice of $k_*$ differently
for each model, choosing $k_*=1.5/\shor$ where 
$\shor=\int c_s d\eta$ is the sound horizon at decoupling
(here $c_s$ is the sound speed and $\eta$ denotes conformal time).
In $\l$-space, this value of $k_*$ corresponds approximately
to the place were we did the splitting in \cite{10par}, \ie, to
the early rise of the first acoustic peak ($\l=100$ for flat 
models, higher/lower $\l$ for open/closed models).

\begin{figure}[tb] 
\centerline{\epsfxsize=9.5cm\epsffile{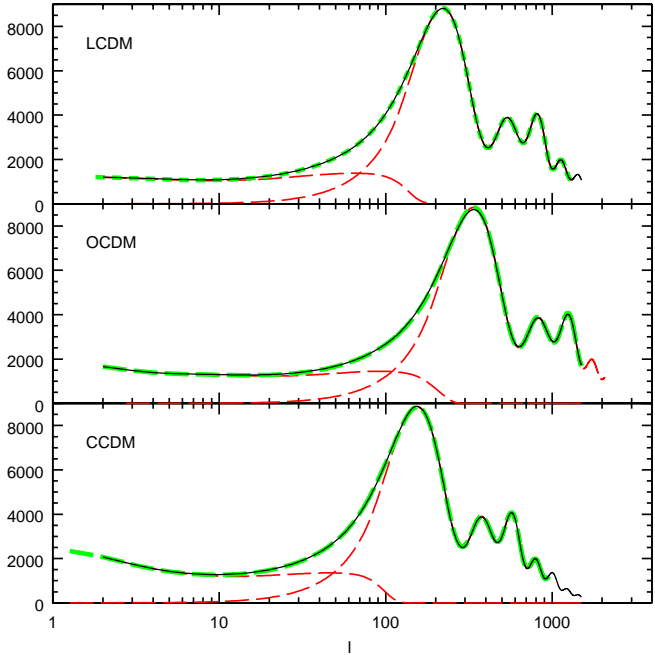}}
\caption{\label{clksplit}\footnotesize%
Three examples of our method for computing CMB power spectra.
The three panels show a
flat $\Lambda$-model ($\Omega_k=0,\Omega_\Lambda=0.7$ --- top),
a moderately open model ($\Omega_k=0.4,\Omega_\Lambda=0.3$ --- middle) 
and a moderately closed model ($\Omega_k=-0.2,\Omega_\Lambda=0.9$ --- bottom). 
All three models have $\omega_{dm}=0.1225$, $\omega_{b}=0.0245$ and
$f_\nu=0$. 
The model $\Chi$ used in the $k>k_*$ calculation
had
$\Omega_k=\Omega_\Lambda=0$, and is simply shifted sideways
differently in each panel, whereas the $(k<k_*)$ spectra $(\Clo)$ 
were computed separately for each model.
The solid line shows the full CMBfast
calculation while the long dashed line shows the result of our new
method, i.e., the sum of the two dashed curves.}
\end{figure}

\Fig{clksplit} shows an example of this splitting technique.
We show three panels with a flat $\Lambda$ model, a moderately open model and 
a moderately closed models. Each panel has four curves. 
The solid line is the model calculated
fully with CMBfast from scratch. The long dashed line shows the same
model calculated with our new technique, and is seen to differ by less than
$2\%$ across the spectrum. For completeness we also show the spectra
for $k<k_*$ and $k>k_*$ that were added in each panel. Note that the $k>k_*$
curve is the same for the three panels, merely shifted sideways by different
amounts to
match the angular diameter distance. 

We do not make a sharp cut at $k_*$. Rather,
to avoid numerical problems, we use a soft cut defined by the
function
\beq{kwin}
w(k)\equiv {2 \over 1+e^{2 \left(k\over k_*\right)^4}},
\eeq 
Specifically, when computing $\Clo$ and $\Chi$, we multiply
the primordial $k$ power spectrum by $w(k)$ and $[1-w(k)]$, 
respectively.
Note that even a sharp $k$-cut would result in a fuzzy $\l$-cut,
since projection effects alias a given $k$-value onto a range of
$\l$-values.

\subsection{Testing the $\Cl$ accuracy}

\begin{figure}[tb] 
\centerline{\epsfxsize=9.5cm\epsffile{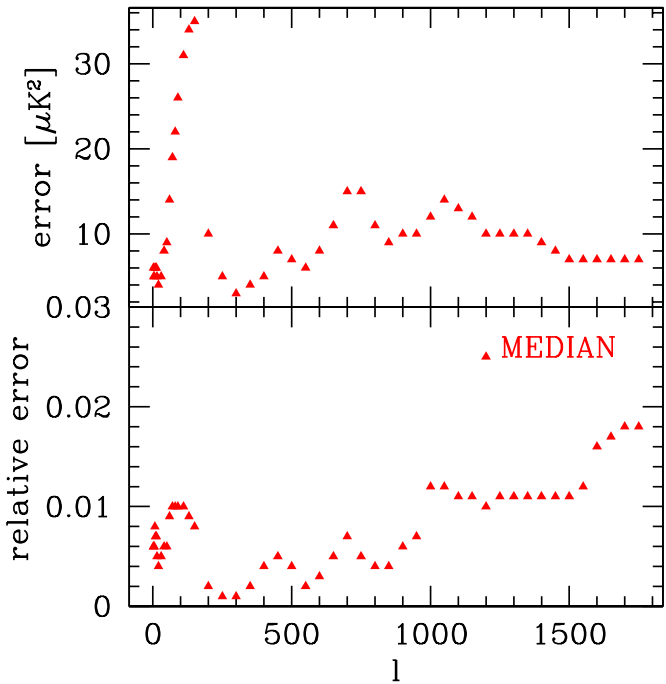}}
\bigskip
\caption{\label{diffcl}\footnotesize%
Median differences between the results of directly computing a model with
CMBfast and the results in our grid as a function of $\l$. 
The top panel shows absolute differences while the lower panel shows
relative errors.}
\end{figure}

To test the accuracy of our method,
we drew a random sample of $\sim 10^3$ of 
the models from our final grid
and recomputed them from scratch with CMBfast.
We also added about $10^2$ models to the test sample by hand 
that we suspected might be particularly troublesome.
The results are shown in \fig{diffcl}.
As can be seen, the median accuracy is better than
$2\%$ for most $\l$-values. It should be noted that this is a test 
not only of the $k$-space splitting technique, but
of our full pipeline, which includes several steps of interpolation.

One interesting thing to point out is that 
our median error is substantially lower than our
mean error. This occurs because there are a small number of 
outlier models where we find significantly larger errors. We should
first note that our worst model is never more that $30\%$ off for
$\l<1000$. An examination of the worst models reveals our main sources
of inaccuracies. Our main source is percent level errors is the
calculation of the angular diameter distances, which lead to a small
relative shift between the spectra. In models with sharp peaks,
this can lead to large relative errors although the two curves are
very similar to each other. Thus this is quite a benign error that can be
further improved by a better calculation of the angular diameter
distance but that has no effect for the likelihood 
of current data which has window functions that are much wider than this
small shift.

The second source of error is in the region were we combine the high
and low $k$ spectra. Usually our worst models (which
are off by less that $15\%$ in this region) have a significant ISW
contribution at low $\l$, so that this ISW contribution is still significant
for the high $k$ wavelengths but were not included because we always 
shift flat $\Omega_m=1$ models. Although this source of error is
inherent to our method, it should not be a source of concern
in practice because
models with such a large ISW contribution
are already ruled out by the data; in particular, they are
inconsistent with the rather flat low $\l$ power spectrum
seen by COBE.

Finally there were errors that could be traced to the coarseness of
our grid. Specifically our low grid had only three values of
$\tau$. Interpolation errors lead to differences (around $10\%$ for
our worst models) for some high $\tau$
models. These errors could be trivially reduced 
by refining our $\tau$-grid although the high $\tau$ models that are
inaccurate are disfavored by the data.

\addtocounter{section}{1}
\smallskip
\centerline{\bf APPENDIX C: METHOD FOR COMPUTING $P(k)$}

\subsection{The approximation}

When normalized according to \eq{PNormalizationEq}, 
the matter power spectrum $P(k)$ depends on the six
parameters $(\Ok,\Ol,\od,\ob,\fn,\ns)$.
However, this dependence can be approximately factored as 
\beqa{PapproxEq}
&&P(k;\Ok,\Ol,\od,\ob,\fn,\ns) \approx\nonumber\\
&&\quad G(\Ok,\Ol)^2 T(hk;\od,\ob,\fn)^2 \left({k\over k_0}\right)^{\ns},
\eeqa
where $h$ is given by \eq{hEq}.
Here $G$ is the growth factor from linear perturbation
theory \cite{Peebles80,Lahav91,Hamilton00} and $T$ is the transfer function normalized 
so that $T = 1$ for $k=0$.
We use $k_0=0.05\,\Mpc^{-1}$ to match the tilt convention of
CMBfast. It is important to 
note that the wave number entering in $T$ is measured using physical distance
units $\Mpc^{-1}$ whereas that entering in $P$ is measured in 
astronomical distance units ($h\,\Mpc^{-1})$.
The approximation given by \eq{PapproxEq} 
becomes exact for the case $\fn=0$, and we will 
quantify its accuracy for the neutrino case below.

We compute $G$ numerically using the publicly available 
Grow$\lambda$ package \cite{Hamilton00}. 
There are excellent packages 
of fitting formulae available for rapid computation 
of the transfer function $T$ \cite{Eisenstein99,Novosyadlyj99},
but unfortunately
none of these currently handle the general case that we need.
Since $T(k)$ depends on merely three parameters, CPU time is not an issue
and we simply compute it numerically using 
CMBfast \cite{cmbfast}.


\subsection{Testing the $P(k)$ accuracy}

\begin{figure}[tb] 
\centerline{\epsfxsize=9.5cm\epsffile{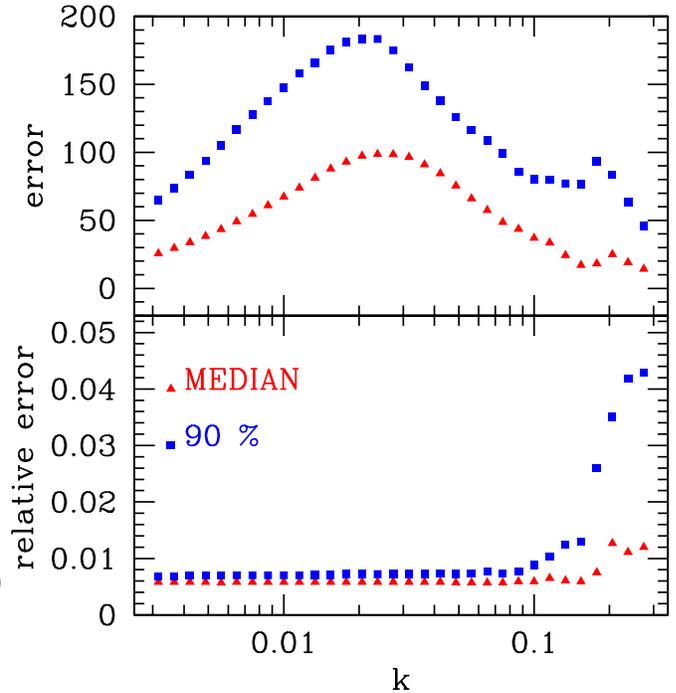}}
\bigskip
\caption{\label{diffpk}\footnotesize%
Differences between the results of directly computing a model with
CMBfast and the results in our grid as a function of wavelength.
The error distribution of each $k$ is illustrated by its
median and 90th percentile.
The top panel shows absolute differences while the lower panel shows
relative errors. 
}
\end{figure}
 
To test the accuracy of our method,
we drew a random sample of $10^3$ of 
the models from our final grid that were not ruled out at more than
$5\sigma$
and recomputed them from scratch with CMBfast.
The results are shown in \fig{diffpk}.
We see that the median accuracy of our
$P(k)$-method is better than $1\%$ for $k\simlt 0.15$
and never gets worse than about $1.4\%$ over our range of interest.
We also tested $10^3$ random models from the full parameter space
(without the $5\sigma$ cut on unphysical models), obtaining median
errors similar to the 90\% curve in \fig{diffpk}.

The only reason that \eq{PapproxEq} is not exact is that
neutrinos affect the growth rate of fluctuations at late
times when $\Ok$ and $\Ol$ become important.
$\fn$ therefore cannot be separated completely from $G$. On the other hand,
$\fn$ cannot be absorbed into $G$ either, since it suppresses 
only small-scale fluctuations.
Fortunately, \fig{diffpk} shows that \eq{PapproxEq} is nonetheless
very accurate in practice, breaking down badly only on
scales smaller than are relevant to our present analysis.
This is because at the low redshifts where $\Ok$ and $\Ol$ become
important, the neutrino free-streaming scale below which fluctuations
are suppressed is 
below a few $h^{-1}\Mpc$ if $m_\nu>1\eV$.
 
Even for $\fn=0$ when \eq{PapproxEq} is strictly speaking
exact, the practical implementation can cause small inaccuracies.
We found our $0.6\%$ ``noise floor'' seen in 
\fig{diffpk} to be due to the horizontal shifting
of the transfer function given by $h$, which we
accomplished with a cubic spline. This effect is also responsible 
for part of the rise in relative errors towards small scales, where
high baryon models have pronounced wiggles.
It should be possible to eliminate this problem by computing 
the input transfer function grid with more finely
spaced $k$-values, able to oversample all 
baryonic wiggles.




\end{document}